\documentclass{article}

\usepackage{microtype}
\usepackage{graphicx}
\usepackage{subcaption}
\usepackage{booktabs} 

\usepackage{hyperref}

\usepackage[preprint]{icml2026}

\usepackage{amsmath}
\usepackage{amssymb}
\usepackage{mathtools}
\usepackage{amsthm}

\usepackage[capitalize,noabbrev]{cleveref}

\theoremstyle{plain}

\theoremstyle{definition}

\theoremstyle{remark}

\usepackage[textsize=tiny]{todonotes}

\icmltitlerunning{Probing 3D Chromatin Structure Awareness in Evo2 DNA Language Model}

\begin{document}

\twocolumn[
  \icmltitle{Probing 3D Chromatin Structure Awareness in Evo2 DNA Language Model}



  \icmlsetsymbol{equal}{*}

  \begin{icmlauthorlist}
    \icmlauthor{UkJin Lee}{wcgs}
  \end{icmlauthorlist}

  \icmlaffiliation{wcgs}{Molecular Biology Program, Weill Cornell Graduate School of Medical Sciences, New York, NY, USA}

  \icmlcorrespondingauthor{UkJin Lee}{ukl4001@med.cornell.edu}

  \icmlkeywords{DNA language model, Evo2, 3D chromatin structure, 3D chromatin architecture, TAD, loop, CTCF, cohesin, in-silico perturbation, sequence generation}

  \vskip 0.3in
]



\printAffiliationsAndNotice{}  

\begin{abstract}

DNA language models like Evo2 now fit million-token contexts large enough to cover entire TADs, yet whether they learn 3D chromatin structure, a key regulatory layer acting atop primary sequence, remains untested and questionable, given that Evo2's training data includes prokaryotes lacking this structure. We probed Evo2-7B on TAD boundaries and convergent CTCF loops in 1\,Mb windows using two complementary tests: likelihood-based perturbation and sequence generation. Evo2 did not distinguish functional perturbations from matched random controls and failed to reliably generate convergent CTCF loops, recovering TAD boundaries only partially. Together, these results indicate that Evo2 has learned local CTCF grammar but misses higher-order 3D organization, pointing to bidirectional model architectures integrating cell types and 3D contacts, rather than longer contexts, as the path to developing 3D-aware DNA language models.
\end{abstract}

\section{Introduction}

\begin{figure}[ht]
    \centering
    \includegraphics[width=\columnwidth]{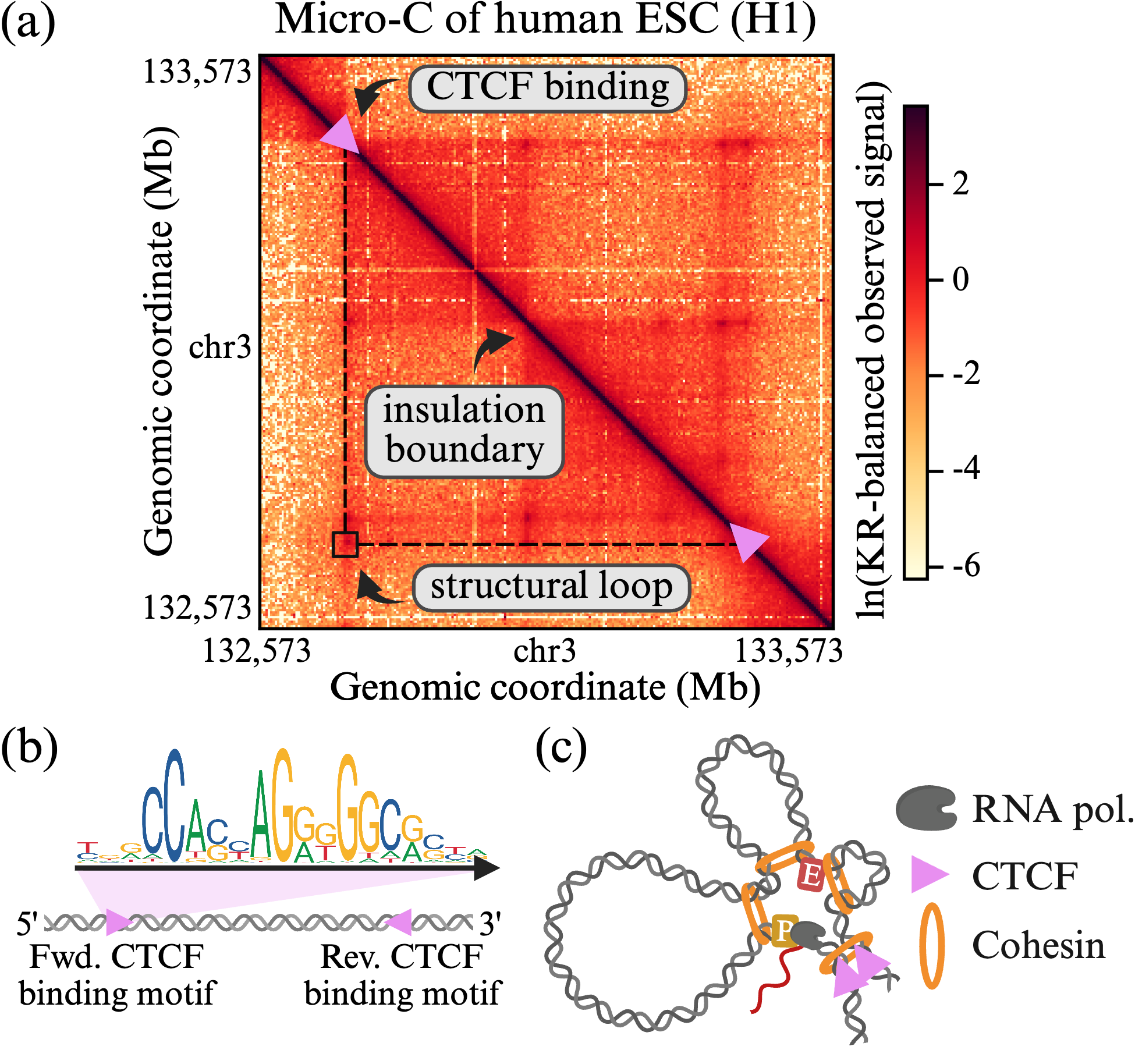}
    \caption{
    \textbf{(a)} Experimental Micro-C data showing TAD (insulation) boundary and convergent CTCF binding forming structural loop in human ESC. \textbf{(b)} Convergent CTCF binding motifs. \textbf{(c)} Illustration showing how CTCF binding and cohesin-mediated chromatin loop extrusion can form insulation boundary and bring distal genomic loci into close proximity in 3D space. Illustration created with biorender.com
    }
    \label{F01}
\end{figure}

Eukaryotic chromosomes are organized into hierarchical 3D structures---compartments, topologically associating domains (TADs), and CTCF/cohesin-mediated loops \cite{Bonev_Cavalli_NatRevGenet2016, Fudenberg_Mirny_CellReports2016} (Fig.~\ref{F01})---that regulate gene expression during development \cite{Furlong_Levine_Science2018} and whose disruption (e.g., TAD boundary deletions \cite{Lupianez_Mundlos_Cell2015} or CTCF inversions \cite{Guo_Wu_Cell2015}) can rewire enhancer--promoter contacts and drive disease \cite{Flavahan_Bernstein_Nature2016}. A DNA language model that reliably encodes 3D chromatin structures could therefore guide synthetic regulatory design, variant interpretation, and structure-preserving genome editing.

Generative DNA language models such as Evo2 \cite{Brixi_Hie_Nature2026}, trained on 9.3 trillion nucleotides with 1-million-token context windows now large enough to span entire TADs, in principle have the capacity to learn this regulatory layer; yet whether they actually do so remains untested, particularly given that Evo2's cross-species training corpus includes prokaryotes that lack the eukaryote-specific CTCF/cohesin machinery. Here we evaluate whether Evo2 implicitly learns 3D chromatin organization via likelihood-based perturbation and sequence generation on TAD boundaries and convergent CTCF loop anchors within 1\,Mb contexts, and find that it neither distinguishes functional 3D elements from matched random controls nor reliably generates 3D-compatible sequences---revealing fundamental limitations of current DNA language models for encoding higher-order genome organization.

\section{Methods}

\subsection{Region Curation}

We curated 1\,Mb windows from hg38 centered on features identified from H1-ESC Micro-C (4DN 4DNES21D8SP8) \cite{Krietenstein_Rando_MolecularCell2020}, CTCF ChIP-seq (ENCODE ENCFF368LWM), and FIMO motif scanning (JASPAR MA0139.1, $p < 10^{-4}$). The \textbf{TAD boundary cohort} ($n=231$) comprised \textbf{strong w/ CTCF} ($n=60$, top-quartile insulation $> 1.034$ overlapping CTCF peaks), \textbf{strong w/o CTCF} ($n=60$), \textbf{weak} ($n=60$), and \textbf{boundary control} ($n=51$, GC-matched, $>$500\,kb from any boundary or CTCF peak, $<$1\% N). The \textbf{structural loop cohort} ($n=120$) comprised convergent CTCF motif pairs ($+$ then $-$) separated by 100\,kb--1\,Mb with experimentally validated Micro-C loops (4DN 4DNFI3RMWQ85).

\subsection{Compute Requirements for Evo2}
We ran Evo2-7B via NVIDIA BioNeMo v2.7: 1\,Mb likelihood scoring required 16 $\times$ A100 80\,GB GPUs (TP\,=\,16); 5\,kb--256\,kb generation required a single A100 80\,GB.

\begin{figure}[ht]
    \centering
    \includegraphics[width=\columnwidth]{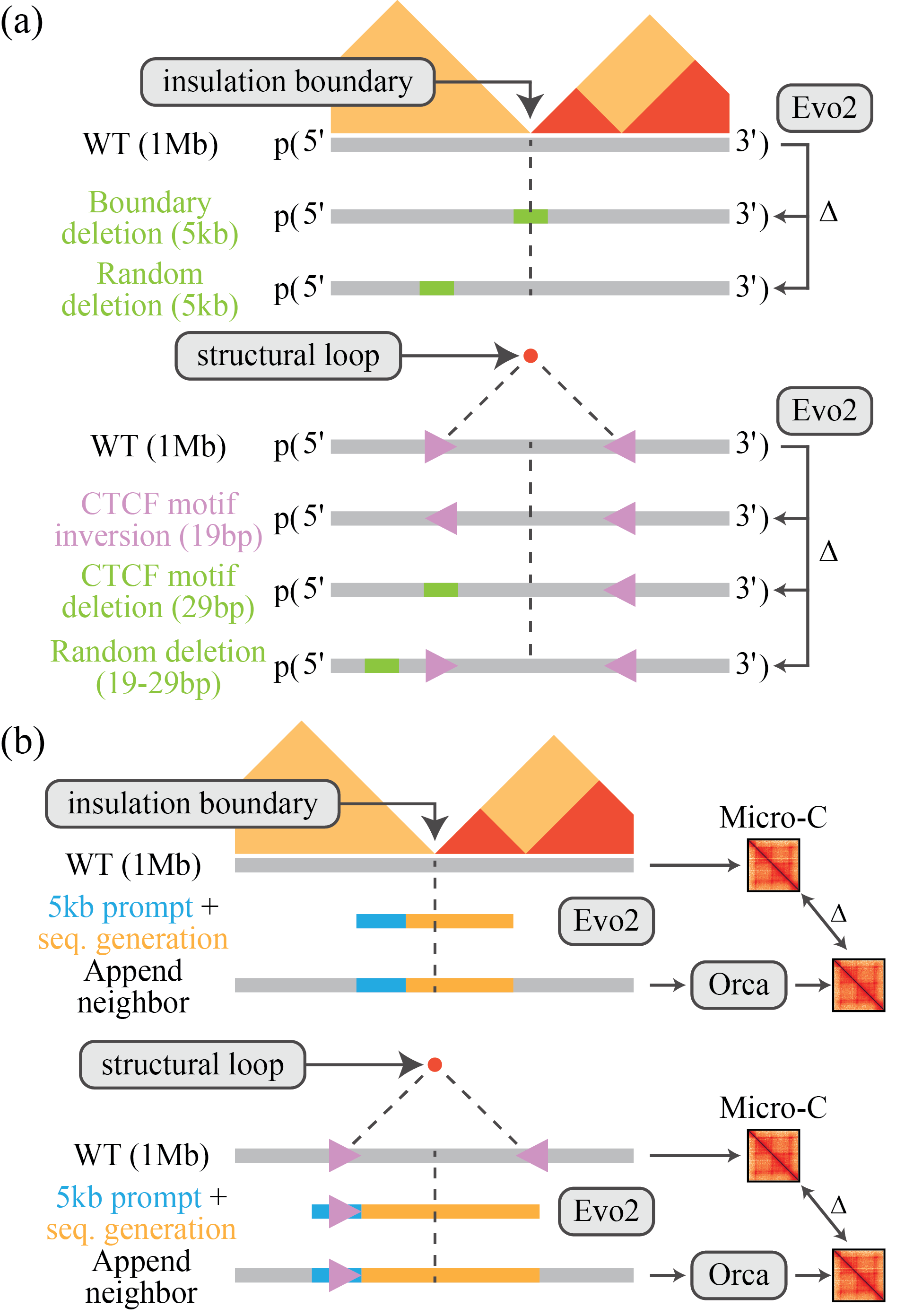}
    \caption{
        \textbf{(a)} Perturbation of insulation boundaries and structural loop anchors used to test Evo2's sensitivity to functional 3D elements.
        \textbf{(b)} Sequence generation pipeline used to evaluate Evo2's ability to generate 3D-compatible sequences.
    }
    \label{F02}
\end{figure}

\subsection{Perturbation Sensitivity Test}
We tested whether Evo2 penalizes functional disruptions of 3D elements more than matched random controls. Each 1\,Mb sequence was scored with Evo2-7B as collapsed mean per-position log-probability $\bar{\mathcal{L}}$, with $\Delta\mathcal{L}_{\text{mean}} = \bar{\mathcal{L}}_{\text{mutant}} - \bar{\mathcal{L}}_{\text{baseline}}$.

\subsubsection{Boundary Deletion Analysis}
From the TAD boundary cohort, a 20\% stratified subsample (12 strong w/ CTCF, 12 strong w/o CTCF, 12 weak, 10 boundary control) yielded 164 sequences: for each of 36 functional regions, one 5\,kb deletion (positions 497,500--502,500 → random nucleotides) and two matched 5\,kb random controls placed $>$50\,kb from any CTCF site or boundary; boundary-control loci received random perturbations only (Fig.~\ref{F02}(a))).

\subsubsection{CTCF Motif Perturbation Analysis}
From the structural loop cohort, a 10\% stratified subsample (12 convergent CTCF pairs) yielded 108 sequences via three perturbation types per region: \textbf{inversion} (in-place reverse-complementation of the 19-bp motif), \textbf{deletion} (19-bp core motif + 5-bp flanks → random nucleotides), and \textbf{matched controls} (size-matched 19 or 29\,bp replacements at positions $>$50\,kb from any CTCF site).

For base-pair-resolution localization, we rescored the 20 strongest-effect deletions and 20 strongest-effect inversions (ranked by $|\Delta\mathcal{L}_{\text{mean}}|$) plus their 40 matched controls.

\subsection{Sequence Generation Test}
We evaluated whether Evo2-generated sequences produce biologically plausible 3D chromatin structure under Orca \cite{Zhou_Zhou_NatGenet2022}, a sequence-to-3D-genome model which was first validated on real sequences from our cohorts against H1-ESC Micro-C (4DNES21D8SP8). Generation used temperature\,=\,0.8, top\_k\,=\,4, top\_p\,=\,1.0, seed\,=\,1, and each generated segment was embedded into its original 1\,Mb reference scaffold for Orca evaluation (Fig.~\ref{F02}(b)).

\subsubsection{TAD (insulation) Boundary Generation}
Each strong w/ CTCF locus received a 32\,kb prompt (positions 465,500--497,500); Evo2 generated a 5\,kb boundary-spanning segment (positions 497,500--502,500). From the Orca-predicted 4\,kb contact map, we computed insulation scores via the diamond method and compared the strongest local minimum near the designed boundary center to the matched real-reference profile.

\subsubsection{Convergent CTCF Loop Generation}
Each loop locus received a 5\,kb upstream prompt containing the forward-strand CTCF motif. Evo2 generated the downstream reverse-strand motif plus a variable-length spacer. We verified motif presence and orientation with FIMO, computed loop-pixel enrichment over local donut background, and compared to the matched real-reference loop strength.

\section{Results}

\subsection{Perturbation Sensitivity Test}
\subsubsection{Boundary Deletions Are Not Penalized More Than Matched Controls}

\begin{figure}[ht]
    \centering
    \includegraphics[width=\columnwidth]{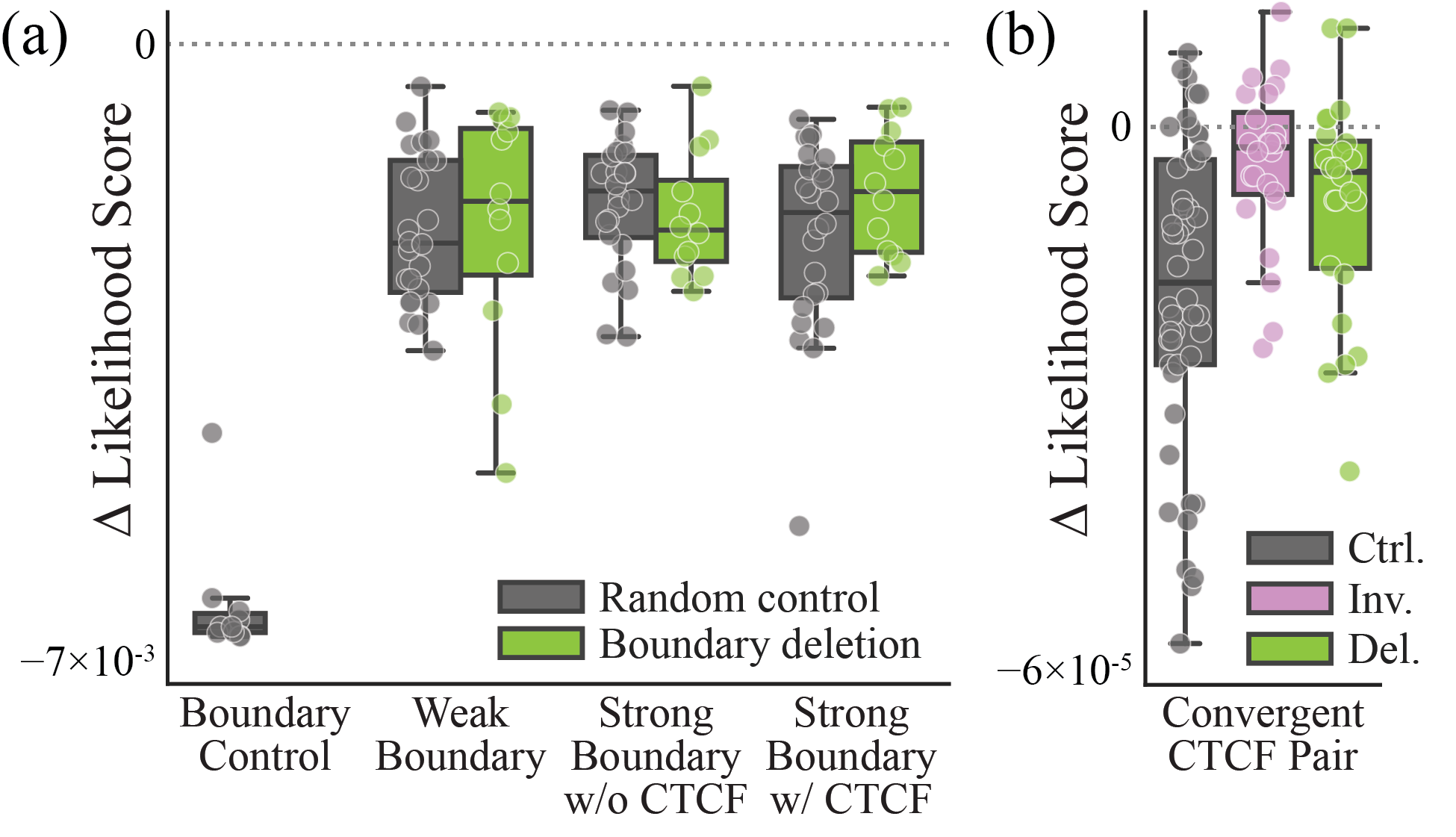}
    \caption{
    \textbf{(a)} $\Delta\mathcal{L}_{\text{mean}}$ for boundary deletions (green, $n=36$) vs.\ matched controls (grey, $n=72$); more negative = stronger penalty. Paired Wilcoxon: strong w/ CTCF $p = 0.204$; strong w/o CTCF $p = 0.519$; weak $p = 0.677$.
    \textbf{(b)} $\Delta\mathcal{L}_{\text{mean}}$ for CTCF inversions (purple, $n=24$), deletions (green, $n=24$), and matched controls (grey, $n=48$); paired Wilcoxon: inversion $p = 0.021$; deletion $p = 0.006$.
    }
    \label{F03}
\end{figure}

Across 36 paired regions, 5\,kb TAD boundary deletions produced \emph{weaker} likelihood penalties than matched random controls (mean paired difference $+1.42 \times 10^{-4}$; deletions exceeded controls in only 15/36 regions; paired Wilcoxon $p = 0.405$), with no category reaching significance (strong w/ CTCF $p = 0.204$; strong w/o CTCF $p = 0.519$; weak $p = 0.677$; Fig.~\ref{F03}(a)). Notably, GC-matched boundary-control loci ($>$500\,kb from any boundary or CTCF peak) produced ${\sim}3\times$ stronger penalties (mean $-6.54 \times 10^{-3}$) than either boundary deletions ($-1.94 \times 10^{-3}$) or their matched controls ($-2.09 \times 10^{-3}$), likely reflecting heterochromatic or Polycomb-repressed sequence grammar that Evo2 has learned more robustly.

\subsubsection{CTCF Motif Edits Are Also Less Penalized Than Matched Controls}

CTCF inversions and deletions were both \emph{less} penalized than their matched random controls (Fig.~\ref{F03}(b)): inversions, mean paired difference $+9.5 \times 10^{-6}$ ($p = 0.021$); deletions, $+1.6 \times 10^{-5}$ ($p = 0.006$). Inversions produced weaker penalties than deletions, suggesting Evo2 has limited sensitivity to the base-pair grammar governing CTCF orientation.

\begin{figure}[ht]
    \centering
\includegraphics[width=\columnwidth]{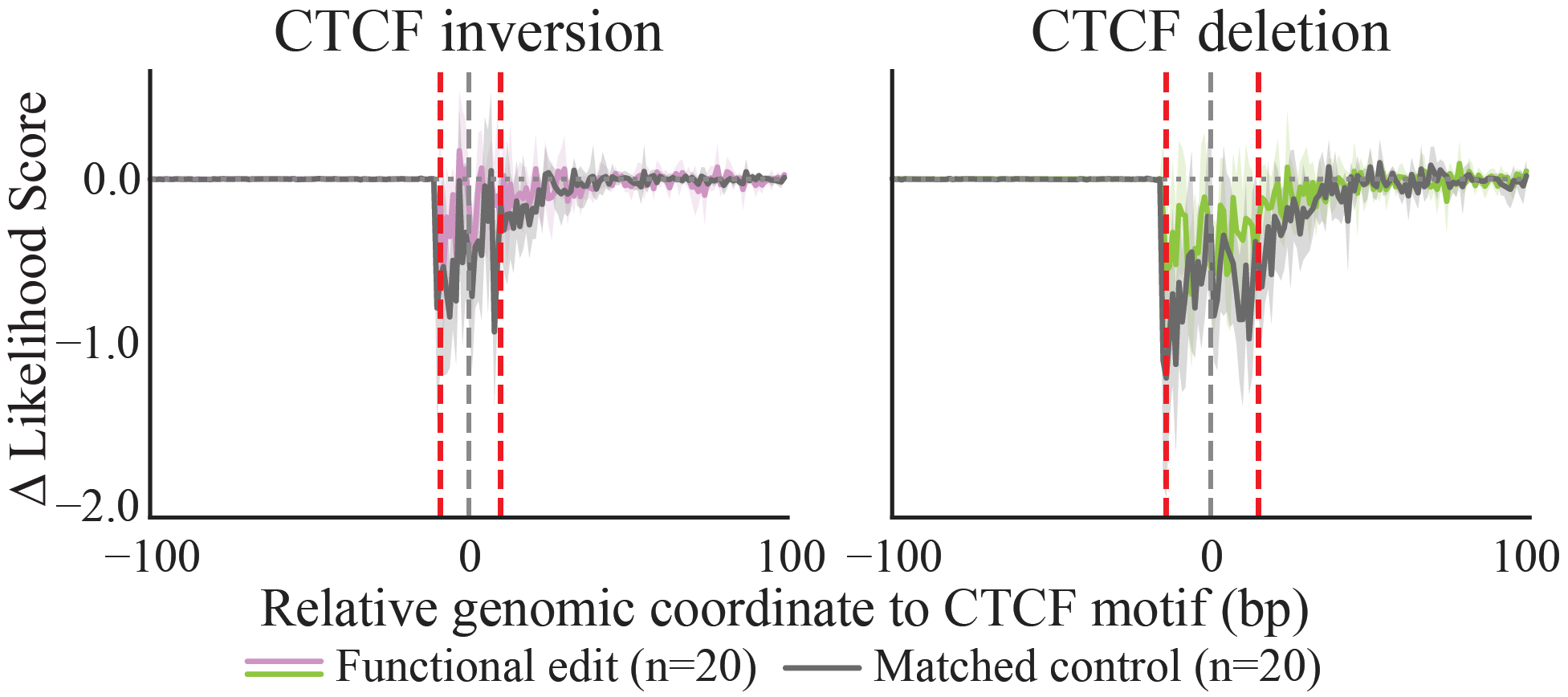}
    \caption{
    Site-centered per-position delta likelihood profiles for \textbf{(a)} CTCF inversions (purple) and \textbf{(b)} deletions (green) vs.\ matched controls (gray); $n=20$ per group. Red dotted lines: edited regions; ribbons: bootstrap 95\% CI.}
    \label{F04}
\end{figure}

Per-position rescoring of the 80-mutant subset revealed that penalties are focal and strictly downstream-biased: signal concentrated within ${\sim}100$--200\,bp 3$'$ of the edit and returned to baseline beyond, consistent with Evo2's autoregressive left-to-right scoring, which alters only downstream prefix context (Fig.~\ref{F04}). At the edited spans themselves, functional inversions ($-0.16$) and deletions ($-0.37$) were again weaker than their matched controls ($-0.46$ and $-0.64$).

\subsection{Sequence Generation Test}

\subsubsection{TAD Boundary Generation Achieves Partial, Heterogeneous Recovery}

\begin{figure}[ht]
    \centering
    \includegraphics[width=\columnwidth]{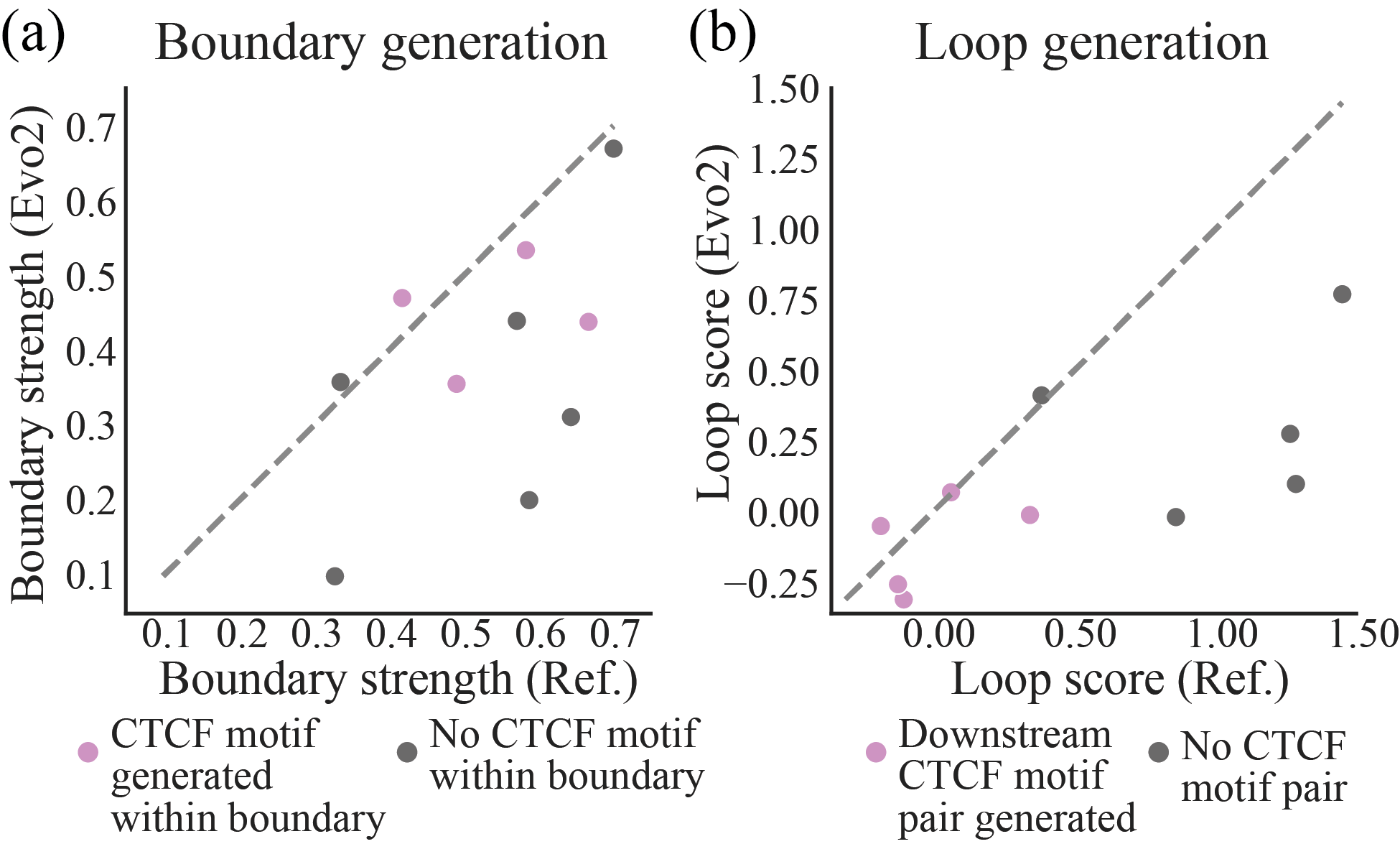}
    \caption{
        Generated vs.\ reference structural scores. \textbf{(a)} TAD boundary insulation score; purple = CTCF motif detected in generated 5\,kb window. \textbf{(b)} CTCF loop enrichment; purple = convergent CTCF motif pair detected. Dashed line: identity.}
    \label{F05}
\end{figure}

\begin{figure}[ht]
    \centering
    \includegraphics[width=\columnwidth]{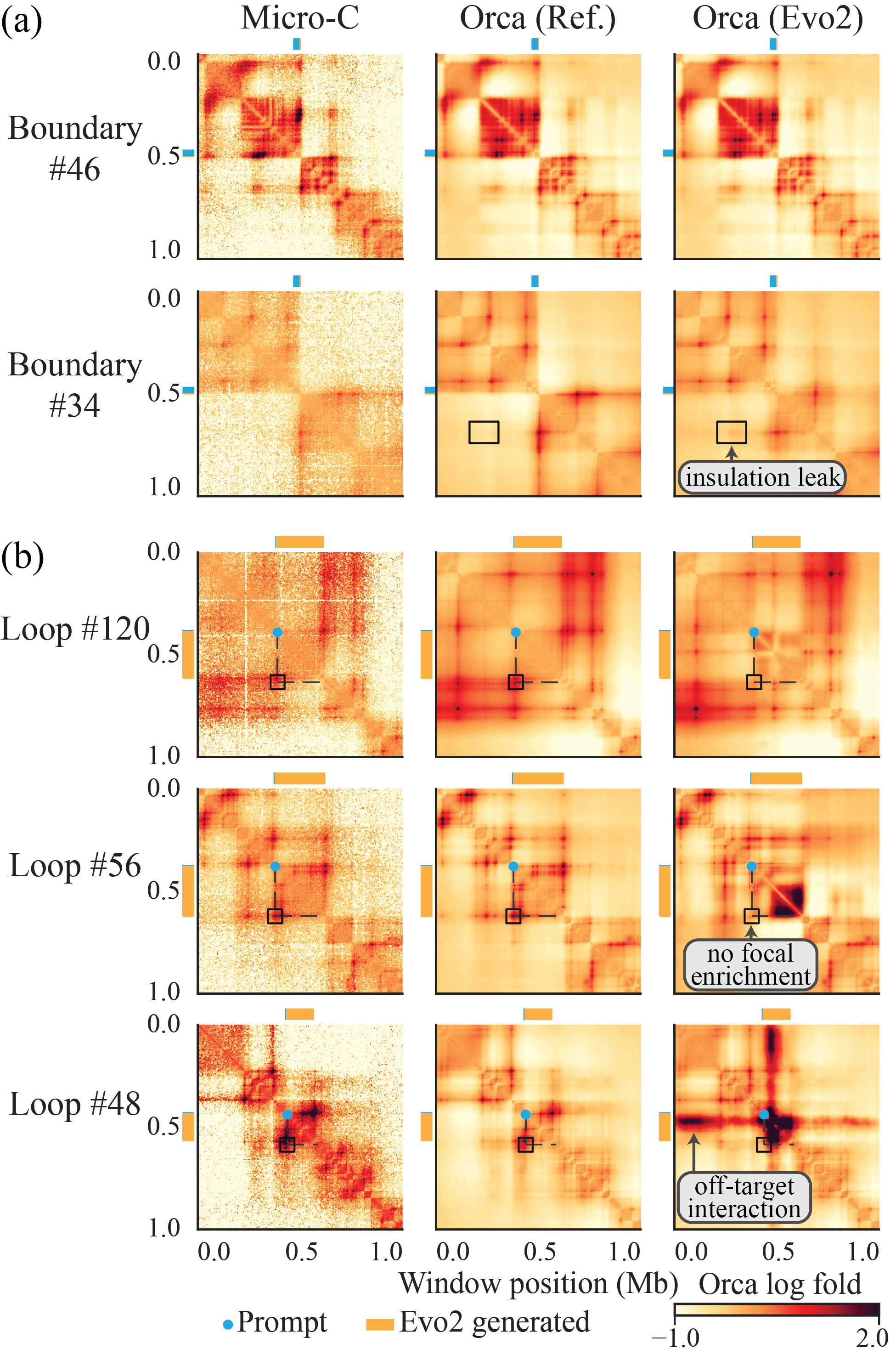}
    \caption{
        Examples of \textbf{(a)} boundary and \textbf{(b)} loop generation. Columns: experimental Micro-C, Orca prediction from reference, Orca prediction from Evo2-generated sequence. Blue: prompt; yellow: generated.
    }
    \label{F06}
\end{figure}

With extensive flanking context (497.5\,kb on each side), Evo2-generated 5\,kb boundary segments showed partial structural recovery: median generated insulation score 0.407 vs.\ reference 0.587 (median delta $-0.134$; Fig.~\ref{F05}(a)). 4 of 10 candidates produced CTCF motifs within the generated window, indicating that Evo2 has learned local CTCF--boundary associations when given sufficient prompt context. Single-locus examples (Fig.~\ref{F06}(a)) show heterogeneous fidelity, with top performers reproducing clear insulation while weaker candidates leak signal across the boundary.

\subsubsection{Loop Generation Fails to Coordinate Convergent Anchors Across Long Distances}

With a minimal 5\,kb prompt containing the upstream forward-strand CTCF motif, Evo2 was tasked with generating the downstream reverse-strand anchor plus intervening spacer (median 175\,kb). Only 5 of 10 candidates produced accepted convergent motif pairs in expected windows (Fig.~\ref{F05}(b)), and generated sequences failed to produce focal loop-like contact enrichment: median generated enrichment 0.054 vs.\ reference 0.388 (median delta $-0.280$; Fig.~\ref{F06}(b)).

Notably, generated sequences still produced strong \emph{local} insulation around individual CTCF anchors despite the short prompt, indicating that Evo2 has learned local CTCF--insulation associations. However, the focal contact enrichment between convergent pairs---the hallmark of cohesin-mediated loop extrusion---was absent in most candidates, revealing a clear limitation in generating sequences that encode coordinated long-range chromatin interactions.

\section{Discussion}
A central question as sequence-based DNA foundation models scale is whether data and context length alone can encode the long-range 3D regulatory logic of eukaryotic genomes \cite{Tiwari_Leslie_NatRevGenet2025a}. We tested this directly against the most representative feature of mammalian 3D organization, CTCF/cohesin-mediated TAD boundaries and convergent CTCF loops, and found that Evo2 captures some local grammar (placing CTCF motifs and reproducing local insulation given sufficient flanking context) but does not penalize disruption of functional elements and largely fails to coordinate convergent anchors across hundreds of kilobases.

Three properties of Evo2's architecture and training likely contribute. Its autoregressive left-to-right scoring breaks the symmetry of double-stranded DNA, so penalties propagate only downstream and orientation-dependent features are intrinsically hard to score; bidirectional, reverse-complement-equivariant architectures such as Caduceus \cite{Schiff_Kuleshov_2024a} and non-autoregressive frameworks such as DNA-Diffusion \cite{DaSilva_Pinello_NatGenet2026a} explicitly address this asymmetry. Its cross-species corpus also includes prokaryotes lacking CTCF/cohesin machinery, diluting eukaryote-specific signal. Most fundamentally, sequence alone is an incomplete substrate because chromatin conformation and its gene regulatory mechanism depends on cell types \cite{Lee_Apostolou_2025}, and a model blind to cell type, or disease context cannot in principle learn such logic. Earlier work has suggested several directions: GraphReg uses Hi-C contacts via Graph Attention Networks \cite{Karbalayghareh_Leslie_GenomeRes.2022a}, EpiGePT conditions on cell-type-specific epigenomic context \cite{Gao_Wong_GenomeBiol2024a}, and long-range benchmarks highlight the remaining challenges \cite{Cheng_Ma_NatCommun2025}. These suggest that 3D-aware DNA language models will require bidirectional architectures, cell type conditioning, and explicit 3D contact inputs rather than longer contexts alone.

Compute cost restricted us to subsampled regions and precluded testing Evo2-40B, and we evaluated only CTCF/cohesin-mediated structures, leaving Polycomb domains, promoter--enhancer hubs, and tissue-specific super-enhancer contacts for future work. The consistency of our results across structural categories nonetheless suggests these limitations are properties of the model itself and offer concrete guidance for future DNA foundation models that reason about the three-dimensional, context-dependent regulatory genome.

\section*{Code Availability}
Pipelines and scripts used for analysis are available at \url{https://github.com/ukjinlee101/evo2-3d-chromatin}.

\section*{Impact Statement}
This paper benchmarks an existing DNA language model (Evo2) on 3D chromatin organization and identifies architectural limitations that should inform the design of future genomic foundation models. By documenting where current sequence-only models fall short of capturing higher-order regulatory structure, our work aims to steer the field toward architectures with more faithful biological inductive biases, which is important for downstream applications in variant interpretation and synthetic regulatory design. We do not release new generative models or sequences intended for biological deployment, and we see no immediate dual-use or societal risks beyond those generally associated with advancing machine learning for genomics.

\section*{Acknowledgements}
We thank Effie Apostolou and the Apostolou lab at Weill Cornell Medicine, and Christina Leslie and the Leslie lab at Memorial Sloan Kettering Cancer Center for helpful discussions. We additionally thank the Leslie lab for providing computational resources.

\nocite{}
\bibliography{references}
\bibliographystyle{icml2026}




\end{document}